# Performance of numerical approximation on the calculation of two-center two-electron integrals over non-integer Slater-type orbitals using elliptical coordinates

A. Bağcı* and P. E. Hoggan

*Institute Pascal, UMR 6602 CNRS, University Blaise Pascal, 24 avenue des Landais BP 80026, 63177 Aubiere Cedex, France*

*albagci@univ-bpclermont.fr*

**Abstract**

The two-center two-electron Coulomb and hybrid integrals arising in relativistic and non-relativistic ab-initio calculations of molecules are evaluated over the non-integer Slater-type orbitals via ellipsoidal coordinates. These integrals are expressed through new molecular auxiliary functions and calculated with numerical Global-adaptive method according to parameters of non-integer Slater-type orbitals. The convergence properties of new molecular auxiliary functions are investigated and the results obtained are compared with results found in the literature. The comparison for two-center two-electron integrals is made with results obtained from one-center expansions by translation of wave-function to same center with integer principal quantum number and results obtained from the Cuba numerical integration algorithm, respectively. The procedures discussed in this work are capable of yielding highly accurate two-center two-electron integrals for all ranges of orbital parameters.

*Correspondence should be addressed to A. Bağcı; e-mail: albagci@univ-bpclermont.fr

## 1. Introduction

The idea of considering the principal quantum numbers over Slater-type orbitals (STOs) in the set of positive real numbers firstly introduced by Parr [1] and performed for the He atom and single-center calculations on the $H_2$ molecule to demonstrate that improved accuracy can be achieved in Hartree-Fock-Roothaan (HFR) calculations. This basis is essential for use in solving the Dirac equation hence the importance of the formalism. Non-integer Slater-type orbitals (NSTOs) are written:

$$\chi_{nlm}(\zeta,\vec{r})=\mathcal{N}_n^S(\zeta)r^{n-1}e^{-\zeta r}Y_{lm}(\theta,\varphi), \tag{1}$$

where,

$$Y_{lm}(\theta,\varphi)=\begin{cases}\frac{1}{2\pi}P_{l|m|}(\cos\theta)e^{im\varphi}\\ \frac{1}{\left[\pi(1+\delta_{m0})\right]^{1/2}}P_{l|m|}(\cos\theta)\begin{cases}\cos|m|\varphi & \text{for} \quad m\geq 0\\ \sin|m|\varphi & \text{for} \quad m<0\end{cases}\end{cases} \tag{2}$$

are complex [2, 3] or real spherical harmonics [4, 5] respectively,

$$\mathcal{N}_n^S(\zeta)=\frac{(2\zeta)^{n+1/2}}{\sqrt{\Gamma(2n+1)}}, \tag{3}$$

are normalization factors, $\Gamma(z)$ are gamma functions. The complexity of molecular integrals over NSTOs have been an important barrier to the implementation of this idea for many years. First attempts to derive analytical relations of molecular integrals over NSTOs were made by Silverstone [6, 7], Geller [8], Bishop and LeClerc [9] at the same time. The remarkable results obtained in atomic properties [10-16] have caused increased popularity of the use of NSTOs over the last decade. Unfortunately, no detailed investigation of molecular electronic structure calculations over NSTOs has been made yet because of ongoing difficulties on the evaluation of multi-center molecular integrals and their lack of the precision.

The approaches used hitherto for evaluation of molecular integrals over NSTOs are based on the one-center expansion which is an expansion of NSTOs with non-integer principal quantum number in terms of STOs orbitals with integer principal quantum numbers [17],

$$\Psi_{nlm}(\zeta,\vec{r})=\lim_{N_e\to\infty}\sum_{\mu=l+1}^{N_e}V_{nl,\mu l}^{*N_e}\Psi_{\mu lm}(\zeta,\vec{r}) \tag{4}$$

here, $n\in\mathbb{R}^+,\mu\in\mathbb{N}^+$ (please see Ref. [18] for definition of $V_{nl,\mu l}^{N_e}$ expansion coefficients), or binomial expansions [19],

$$(a+b)^{N_1}(a-b)^{N_2}=\lim_{N_s\to\infty}\sum_{s=0}^{N'_1+N_s} F_s^{N_s}(N_1,N_2)\,a^{N_1+N_2-s}\,b^s, \tag{5}$$

where, $N'_1$ is the integer part of principal quantum number $N_1$. Detailed investigations of both methods have been made recently for two-center overlap integrals in [20]. The results presented in [20] show that these suggested methods are unable to calculate molecular integrals over NSTOs accurately. Evaluation of two-center overlap integrals over NSTOs via ellipsoidal coordinates through Eq. (5) with an infinite series shows it is ill-conditioned, making it necessary to take into consideration thousands of terms. Besides, the results given in [20] become numerical proof to discussions on the one-center expansion method made in [21, 22].

The NSTOs provide a more flexible basis for molecular calculation then STOs. Nevertheless, it may not be said they have vital importance in non-relativistic electronic structure theory. Many calculations [23-28] show that obtaining the adequate level of accuracy is possible by using appropriate basis sets of STOs. In fact, the NSTOs start to play a major role in relativistic electronic structure theory of molecules and this increases importance of accurate calculation of molecular integrals over NSTOs.

The Dirac spectrum can be altered by negative-energy contributions, in relativistic electronic structure calculation the whole spectrum is needed for mathematical completeness and contribution of the negative-energy states can significantly improve the accuracy [29]. The variational instability, which is referred to as variational collapse [30, 31], may arise throughout the solution of the Dirac equation to find the lowest eigenvalues of positive-energy states due to the Dirac Hamiltonian being unbounded from below [32]. The choice of basis functions, which also should provide the conditions specified for the non-relativistic approach [33, 34], depends on the kinetic balance condition [35-37],

$$f^{\beta}_{n_r\kappa}(\zeta,\vec{r})=\left(\beta\frac{d}{dr}+\frac{\kappa}{r}\right)f^{-\beta}_{n_r\kappa}(\zeta,\vec{r})\,;\;\; c\to 0 \tag{6}$$

where, $c$ is the speed of light, $\beta=\pm 1$, and represent large and small components of the wave-function, $-|\kappa|+\frac{1}{2}\leq\mu\leq|\kappa|-\frac{1}{2}$, $n_r$ are radial quantum numbers [38], $\zeta$ are orbital parameters, guarantees the separation of positive- and negative-energy states.

On account of the kinetic balance condition the exponential-type orbitals (ETOs) to be chosen in relativistic electronic structure calculations should be a limit or reduced case of L-Spinors [39, 40, 41], which are complete square-integrable basis functions in terms of generalized Laguerre polynomials,

$$f^{\beta}_{n_r\kappa\mu}(\zeta,\vec{r})=\mathcal{N}^{L}_{n_r\kappa}(2\zeta r)^{\gamma-1}e^{-\zeta r}\left\{-(1-\delta_{n_r 0})L^{2\gamma}_{n_r-1}(2\zeta r)+\beta\left(\frac{N_{n_r\kappa}-\kappa}{n_r+2\gamma}\right)L^{2\gamma}_{n_r}(2\zeta r)\right\}\Omega^{\beta}_{\kappa\mu}(\theta,\varphi), \tag{7}$$

here, $\gamma=\sqrt{\kappa^2-Z^2/c^2}$ , $N_{n_r\kappa}=\sqrt{n_r^2+2n_r\gamma+\kappa^2}$ and $\mathcal{N}^{L}_{n_r\kappa}$ are the normalization factors obtained by the standard orthogonality properties of Laguerre polynomials,

$$\mathcal{N}^{L}_{n_r\kappa}=\sqrt{\frac{n_r!(2\gamma+n_r)}{2N_{n_r\kappa}(N_{n_r\kappa}-\kappa)\Gamma(2\gamma+n_r)}}, \tag{8}$$

respectively. The $\Omega_{\kappa\mu}(\theta,\varphi)$ in Eq. (7) are the spin $\frac{1}{2}$ spinor spherical harmonics and following [2, 3, 38, 42] are defined as,

$$\Omega^{\beta}_{\kappa\mu}(\theta,\varphi)=\frac{1}{\sqrt{2\beta\kappa+1}}\begin{bmatrix} sgn(-\beta\kappa)\sqrt{(\beta\kappa)+1/2-\mu}\,Y_{l_\beta,\mu-1/2}(\theta,\varphi) \\ \sqrt{(\beta\kappa)+1/2+\mu}\,Y_{l_\beta,\mu+1/2}(\theta,\varphi)\end{bmatrix}, \tag{9}$$

where,

$$l_\beta=\begin{cases}\beta\kappa & \beta\kappa>0\\ -(\beta\kappa)-1 & \beta\kappa<0\end{cases}$$

and, $Y_{lm}(\theta,\varphi)$ are the complex spherical harmonics. Note that, the operator $\hat{K}_{\theta,\varphi}$,

$$\hat{K}_{\theta,\varphi}\equiv(\hat{\sigma}.\hat{r})=\begin{bmatrix} Cos(\theta) & Sin(\theta)e^{-i\varphi}\\ Sin(\theta)e^{i\varphi} & -Cos(\theta)\end{bmatrix}, \tag{10}$$

with, $\hat{\sigma}$ are Pauli spin operators, $\hat{r}=\frac{\vec{r}}{|r|}$ , changes the parity of spinor spherical harmonics since it is odd of parity [43],

$$\hat{K}_{\theta,\varphi}\Omega^{\beta}_{\kappa\mu}(\theta,\varphi)=-\Omega^{-\beta}_{\kappa\mu}(\theta,\varphi). \tag{11}$$

The L-spinors are related analogously to the Dirac hydrogenic solutions and they smoothly reduce to Sturmian orbitals in the non-relativistic limit [39, 40, 41]. They are most useful for hydrogenic problems, either for isolated atoms or for atoms in static electromagnetic fields [44]. However, they are not to be used as a basis to construct molecular electronic wave-functions in the Linear Combination of Atomic Orbital (LCAO-MO) method [45] directly. Any kind of relativistic exponential-type orbitals, obtained from the limit or reduced case of L-spinors, is available to be used in LCAO-MO method and these ETOs can be represent by the finite summation of NSTOs with following relation of generalized Laguerre polynomials $(L^{p}_{q})$ [46, 47];

$$L_q^p(x)=\sum_{s=0}^{q}\omega_q^p(s)x^s, \tag{12}$$

$$\omega_q^p(s)=\frac{(-1)^s}{s!}\left(\frac{\Gamma(q+p+1)}{\Gamma(s+p+1)}\times\frac{1}{\Gamma(q-s+1)}\right). \tag{13}$$

The accurate evaluation of two-center electron repulsion integrals over NSTOs constitute the basic building block in the study of relativistic molecular systems and they arise not only in their own right for the calculation of diatomic molecules, but they are also central to the calculation of the multi-center multi-electron integrals.

In this paper, new molecular auxiliary functions ${}^{P_i,Q_i}\mathcal{G}^{N_1,q}$, which are obtained with the generalization of the solution of Poisson equation as a partial differential equation in spherical coordinates by expanding the potential in the new set of functions, referred to as spectral forms (SFs) [48, 49] have been introduced. The recurrence relations of these auxiliary functions are also obtained. The analytically closed form expressions of the two-center two-electron Coulomb and hybrid integrals over NSTOs are derived. Afterwards, the highly accurate calculation of Coulomb and hybrid integrals over STOs and NSTOs using numerical Global-adaptive method by specify Gauss-Kronrod rule (Please see [20] for details) is performed. A computer program is constructed in the Mathematica programming language [50] with the included numerical methods, the calculations are performed for arbitrary values of quantum numbers and orbital parameters. The results obtained are compared with the results those found in the literature in the case of STOs. The comparison for NSTOs have been made with results obtained from one-center expansion method [17, 18] and Cuba numerical integration library with cuhre algorithm [51, 52]. The benchmark results are presented in the given tables with 25 accurate-decimals.

Note that, the methods proposed here are essentially based on three papers; one of them provides an efficient way to represent the electron repulsion integrals in ellipsoidal coordinates according to a proposed method for the product of angular momentum functions [53], in another, suggested spectral forms expansions allow the radial functions occurring in one-center potential in the case of non-integer values of quantum numbers to be expressed analytically [48], and, highly accurate calculation of these integrals becomes possible by the making use numerical Global-adaptive procedure proposed in last one [20].

## 2. Two-center Coulomb and Hybrid Integrals

The expressions for two-center two-electron Coulomb energy, which enables the potential to be evaluated from spectral forms of Poisson's equation [48] associated with charge density $\rho(\vec{r})$ written as,

$$E=-\frac{1}{4\pi}\int V\left(\vec{r}_{a_2}\right)\nabla^2_{\vec{r}_{b_2}}V^*\left(\vec{r}_{b_2}\right)dV_2=\int V\left(\vec{r}_{a_2}\right)\rho\left(\vec{r}_{b_2}\right)dV_2, \tag{14}$$

where, $dV=r^2\sin(\theta)d\theta d\varphi$, the range of integrals $[0,\infty)\times[0,\pi]\times[0, 2\mathrm{pi}]$. Note that these expressions are symmetric with respect to exchange in subscripts $a,b$.

Following a similar procedure to that given in [48, 49] the radial functions are obtained by expanding the one-center potential,

$$V\left(r_{a_2}\right)=\int\frac{\rho\left(\vec{r}_{a_1}\right)}{r_{12}}dV_1, \tag{15}$$

taking into consideration,

$$\frac{1}{r_{12}}=\sum_{l=0}^{\infty}\sum_{m=-l}^{l}\left(\frac{4\pi}{2l+1}\right)\frac{r_<^l}{r_>^{l+1}}S_{lm}(\theta_1,\varphi_1)S^*_{lm}(\theta_2,\varphi_2) \tag{16}$$

in the new set of function, for normalized NSTOs can be determined as,

$$V(r_{a_2})=\sum_{L_1M_1}F^{L_1}_{N_1}\left(x_1,\vec{r}_{a_2}\right)\mathcal{C}_{L_1M_1}, \tag{17}$$

$$F^{L_1}_{N_1}\left(x_1,\vec{r}_{a_2}\right)=\mathcal{N}_{n_1n_1'}\left(1,t_1\right)\left(2\xi_1\right)f^{L_1}_{N_1}\left(x_1,r_{a_2}\right)S^*_{L_1M_1}\left(\theta_{a_2}\varphi_{a_2}\right) \tag{18}$$

$$f^{L_1}_{N_1}\left(x_1,r_{a_2}\right)=\Gamma\left(N_1+L_1+1\right)\frac{1}{x_1^{L_1+1}}\{P[N_1+L_1+1,x_1]+\frac{x_1^{2L_1+1}}{(N_1-L_1)_{2L_1+1}}Q[N_1-L_1,x_1]\}, \tag{19}$$

and,

$$\mathcal{C}_{L_1M_1}=\left(\frac{4\pi}{2L_1+1}\right)^{1/2}C^{L_1M_1}(l_1m_1,l_1'm_1')A^{M_1}_{m_1m_1'}, \tag{20}$$

respectively. Here,

$$\mathcal{N}\left(p_i,t_i\right)=\frac{[p_i+t_i]^{n_i+1/2}[p_i-t_i]^{n_i'+1/2}}{\sqrt{\Gamma[2n_i+1]\Gamma[2n_i'+1]}}, \tag{21}$$

$x_i=2\xi_i r_{a_i}$, $\xi_i=\frac{1}{2}\left(\zeta_i+\zeta_i'\right)$ $p_i=\frac{R}{2}\left(\zeta_i+\zeta_i'\right)$, $t_i=\frac{\zeta_i-\zeta_1'}{\zeta_i+\zeta_i'}$, $N_i=n_i+n_i'$ and, $\Gamma[\alpha]$ is the Gamma function, $P[\alpha,x]$ is the normalized incomplete gamma, $Q[\alpha,x]$ is normalized complementary incomplete gamma functions respectively and $C^{L|M|}$ are the generalized Gaunt coefficients. Please see Refs.[53, 54] for the definitions of generalized Gaunt coefficients and $A^M$ coefficients.

The charge densities of NSTOs occurring in Eqs. (14, 15) for the same center $(\rho_{aa})$ and different centers $(\rho_{ab})$ defined as,

$$\begin{aligned}&\rho_{aa}(\zeta,\zeta')=\chi_{nlm}(\zeta,\vec{r}_a)\chi^*_{n'l'm'}(\zeta',\vec{r}_a)=\\&=R_n(\zeta,r)R_{n'}(\zeta,r)S_{lm}(\theta_a,\varphi_a)S^*_{l'm'}(\theta_a,\varphi_a),\end{aligned} \tag{22}$$

with,

$$S_{lm}(\theta_a,\varphi_a)S^*_{l'm'}(\theta_a,\varphi_a)=\sum_{LM}\left(\frac{2L+1}{4\pi}\right)^{1/2}C^{L|M|}(lm,l'm')A^M_{mm'}S^*_{LM}(\theta_a,\varphi_a), \tag{23}$$

$$|l-l'|\le L\le l+l';M=|m\pm m'|,$$

and,

$$\begin{aligned}&\rho_{ab}(\zeta,\zeta')=\chi_{nlm}(\zeta,\vec{r}_a)\chi^*_{n'l'm'}(\zeta',\vec{r}_b)=\\&=R_n(\zeta,r_a)R_{n'}(\zeta',r_b)S_{lm}(\theta_{a,},\varphi_a)S^*_{l'm'}(\theta_b,\varphi_b).\end{aligned} \tag{24}$$

By making use of ellipsoidal coordinates $(\mu,\nu,\varphi)$, where, $[1,\infty)\times[-1,1]\times[0,2\varphi]$, and

$$r_{a_i}=\frac{R}{2}(\mu+\nu);r_{b_i}=\frac{R}{2}(\mu-\nu);\cos\theta_{a_i}=\frac{1+\mu\nu}{\mu+\nu};\cos\theta_{b_i}=\frac{1-\mu\nu}{\mu-\nu}, \tag{25}$$

$$dV=\left(\frac{R}{2}\right)^3(\mu^2-\nu^2)d\mu d\nu d\varphi, \tag{26}$$

the two-center two-electron Coulomb integrals,

$$\begin{aligned}&\mathcal{J}^{aa,bb}_{n_1l_1m_1,n_1'l_1'm_1';n_1l_2m_2,n_2'l_2'm_2'}(\zeta_1,\zeta_1';\zeta_2,\zeta_2')=\\&=\int\chi^*_{n_1l_1m_1}(\zeta_1,\vec{r}_{a_1})\chi_{n_1'l_1'm_1'}(\zeta_1',\vec{r}_{a_1})\chi_{n_2l_2m_2}(\zeta_2,\vec{r}_{b_2})\chi^*_{n_2'l_2'm_2'}(\zeta_2',\vec{r}_{b_2})dV_1dV_2,\end{aligned} \tag{27}$$

in the lined-up coordinate system $(|m_i|=|m_i'|=\lambda;\varphi_{a_i}=\varphi_{b_i}=\varphi)$ considering Eq. (15) with Eqs. (17-23) can be obtained as,

$$\begin{aligned}&\mathcal{J}^{aa,bb}_{n_1l_1m_1,n_1'l_1'm_1';n_1l_2m_2,n_2'l_2'm_2'}(\zeta_1,\zeta_1';\zeta_2,\zeta_2')=\\&=\int\sum_{L_1L_2M}\left(\frac{2L_2+1}{2L_1+1}\right)^{1/2}A^M_{m_1m_1'}A^M_{m_2m_2'}C^{L_1|M|}(l_1m_1;l_1'm_1')C^{L_2|M|}(l_2m_2;l_2'm_2')\times\\&\times f^{L_1}_{N_1}(x_1,r_{a_2})R_{n_2}(\zeta_2,r_{b_2})R_{n_2'}(\zeta_2',r_{b_2})\mathcal{P}_{L_1\lambda}(\cos\theta_{a_2})\mathcal{P}_{L_2\lambda}(\cos\theta_{b_2})dV_1dV_2.\end{aligned} \tag{28}$$

And, the right-hand side of Eq. (14) with Eq. (17) in order to find analytical relations for the hybrid integrals,

$$\mathcal{J}^{aa,ab}_{n_1 l_1 m_1, n_1' l_1' m_1'; n_1 l_2 m_2, n_2' l_2' m_2'}\left(\zeta_1,\zeta_1';\zeta_2,\zeta_2'\right)= \\ =\int \chi^*_{n_1 l_1 m_1}(\zeta_1,\vec{r}_{a_1})\chi_{n_1' l_1' m_1'}(\zeta_1',\vec{r}_{a_1})\chi_{n_2 l_2 m_2}(\zeta_2,\vec{r}_{a_2})\chi^*_{n_2' l_2' m_2'}(\zeta_2',\vec{r}_{b_2})\,dV_1 dV_2 \tag{29}$$

can be re-written as,

$$E=\int V\left(\vec{r}_{a_2}\right)\chi_{n_2 l_2 m_2}(\zeta_2,\vec{r}_{a_2})\chi^*_{n_2' l_2' m_2'}(\zeta_2',\vec{r}_{b_2})\,dV_2. \tag{30}$$

Taking into account Eq. (24), the hybrid integrals are determined by,

$$\mathcal{J}^{aa,ab}_{n_1 l_1 m_1, n_1' l_1' m_1'; n_1 l_2 m_2, n_2' l_2' m_2'}\left(\zeta_1,\zeta_1';\zeta_2,\zeta_2'\right)= \\ =\int \sum_{L_1 M_1 L_2}\left(\frac{2L_2+1}{2L_1+1}\right)^{1/2} A^{M_1}_{m_1 m_1'} A^{m_2'}_{M_1 m_2} C^{L_1|M_1|}(l_1 m_1; l_1' m_1') C^{L_2|m_2'|}(L_1 M_1; l_2 m_2)\times \\ \times f^{L_1}_{N_1}\left(x_1, r_{a_2}\right) R_{n_2}(\zeta_2, r_{a_2}) R_{n_2'}(\zeta_2', r_{b_2}) \mathcal{P}_{L_2\lambda}(\cos\theta_{a_2})\mathcal{P}_{l_2'\lambda}(\cos\theta_{b_2})\,dV_1\,dV_2. \tag{31}$$

Finally, the two-center two-electron Coulomb and hybrid integrals arising from the product of two associated Legendre functions with different centers in ellipsoidal coordinates [53], is written

$$\mathcal{P}_{L_1\lambda}(\cos\theta_a)\mathcal{P}_{L_2\lambda}(\cos\theta_b)=\sum_{\alpha=-\lambda}^{L_1}\sum_{\beta=\lambda}^{L_2}\sum_{q=0}^{\alpha+\beta} g^q_{\alpha\beta}(L_1\lambda, L_2\lambda)\left[\frac{\mu\nu^q}{(\mu+\nu)^\alpha(\mu-\nu)^\beta}\right] \tag{32}$$

and the Coulomb integrals may be written,

$$\mathcal{J}^{aa,bb}_{n_1 l_1 m_1, n_1' l_1' m_1'; n_1 l_2 m_2, n_2' l_2' m_2'}\left(\zeta_1,\zeta_1';\zeta_2,\zeta_2'\right)= \\ =\frac{2}{R}\mathcal{N}_{n_1 n_1'}(1,t_1)\mathcal{N}_{n_2 n_2'}(p_2,t_2)\times \\ \times\sum_{L_1 L_2 M}\left(\frac{2L_2+1}{2L_1+1}\right)^{1/2} A^{M}_{m_1 m_1'} A^{M}_{m_2 m_2'} C^{L_1|M|}(l_1 m_1; l_1' m_1') C^{L_2|M|}(l_2 m_2; l_2' m_2')\times \\ \times\Gamma(n_1+n_1'+L_1+1)\frac{1}{p_1^{L_1}}\sum_{\alpha\beta\gamma} g^q_{\alpha\beta}(L_1\lambda, L_2\lambda)\times \\ \times\left\{{}^{P_1}\mathcal{G}^{0,q}_{L_1+\alpha, n_2+n_2'-\beta-1, n_1+n_1'+L_1+1}(p_1,p_2,-p_2)+{}^{Q_1}\mathcal{G}^{2L_1+1,q}_{\alpha-(L_1+1), n_2+n_2'-\beta-1, n_1+n_1'+L_1+1}(p_1,p_2,-p_2)\right\}, \tag{33}$$

where, $max[|-L_1,-L_2|]\le M\le min[L_1+L_2]$, $|l_1-l_1'|\le L_1\le l_1+l_1'$, $|l_2-l_2'|\le L_2\le l_2+l_2'$,

the hybrid integrals may be written,

$$
\begin{aligned}
&\mathcal{J}^{aa,ab}_{n_1 l_1 m_1, n_1' l_1' m_1'; n_1 l_2 m_2, n_2' l_2' m_2'}\left(\zeta_1, \zeta_1'; \zeta_2, \zeta_2'\right) = \\
&= \frac{2}{R} \mathcal{N}_{n_1 n_1'}(1, t_1) \mathcal{N}_{n_2 n_2'}(p_2, t_2) \times \\
&\times \sum_{L_1 M_1 L_2} \left(\frac{2\mathrm{L}_2+1}{2\mathrm{L}_1+1}\right)^{1/2} A^{M_1}_{m_1 m_1'} A^{m_2'}_{M_1 m_2} C^{L_1 |M_1|}(l_1 m_1; l_1' m_1') C^{L_2 |m_2'|}(L_1 M_1; l_2 m_2) \times \\
&\times \Gamma(n_1 + n_1' + L_1 + 1) \frac{1}{p_1^{L_1}} \sum_{\alpha\beta\gamma} g^q_{\alpha\beta}(L_1 \lambda, L_2 \lambda) \times \\
&\times \left\{ {}^{P_1}\mathcal{G}^{0,q}_{L_1+\alpha+1-n_2, n_2'-\beta, n_1+n_1'+L_1+1}(p_1, p_2, p_2 t_2) + {}^{Q_1}\mathcal{G}^{2\mathrm{L}_1+1,q}_{\alpha-L_1-n_2, n_2'-\beta-1, n_1+n_1'+L_1+1}(p_1, p_2, p_2 t_2) \right\},
\end{aligned} \tag{34}
$$

where, $|l_1 - l_1'| \le L_1 \le l_1 + l_1' \quad -L_1 \le M_1 \le L_1, \quad |L_1 - l_2| \le L_2 \le L_1 + l_2.$

Thus, the auxiliary functions occurring in analytically closed form expressions given in Eqs. (33, 34) are first kinds of ${}^{P_i}\mathcal{G}^{N,q}$ and ${}^{Q_i}\mathcal{G}^{N,q}$ auxiliary functions which given as, for normalized incomplete gamma functions,

$$ {}^{P_1}\mathcal{G}^{N_1,q}_{N_2 N_3 N_4}(p_1, p_2, p_3) = \frac{p_1^{N_1}}{(N_4 - N_1)_{N_1}} \int_0^\infty \int_{-1}^1 (\mu\nu)^q (\mu+\nu)^{N_2} (\mu-\nu)^{N_3} P_1[N_4 - N_1, p_1(\mu+\nu)] e^{-p_2\mu - p_3\nu} d\mu\, d\nu \tag{35} $$

$$ {}^{P_2}\mathcal{G}^{N_1,q}_{N_2 N_3 N_4}(p_1, p_2, p_3) = \frac{p_1^{N_1}}{(N_4 - N_1)_{N_1}} \int_0^\infty \int_{-1}^1 (\mu\nu)^q (\mu+\nu)^{N_2} (\mu-\nu)^{N_3} P_2[N_4 - N_1, p_1(\mu-\nu)] e^{-p_2\mu - p_3\nu} d\mu\, d\nu \tag{36} $$

$$ {}^{P_3}\mathcal{G}^{N_1,q}_{N_2 N_3 N_4}(p_1, p_2, p_3) = \frac{p_1^{N_1}}{(N_4 - N_1)_{N_1}} \int_0^\infty \int_{-1}^1 (\mu\nu)^q (\mu+\nu)^{N_2} (\mu-\nu)^{N_3} P_3[N_4 - N_1, p_1(\mu\nu)] e^{-p_2\mu - p_3\nu} d\mu\, d\nu, \tag{37} $$

and, for normalized complementary incomplete gamma functions,

$$ {}^{Q_1}\mathcal{G}^{N_1,q}_{N_2 N_3 N_4}(p_1, p_2, p_3) = \frac{p_1^{N_1}}{(N_4 - N_1)_{N_1}} \int_0^\infty \int_{-1}^1 (\mu\nu)^q (\mu+\nu)^{N_2} (\mu-\nu)^{N_3} Q_1[N_4 - N_1, p_1(\mu+\nu)] e^{-p_2\mu - p_3\nu} d\mu\, d\nu \tag{38} $$

$$ {}^{Q_2}\mathcal{G}^{N_1,q}_{N_2 N_3 N_4}(p_1, p_2, p_3) = \frac{p_1^{N_1}}{(N_4 - N_1)_{N_1}} \int_0^\infty \int_{-1}^1 (\mu\nu)^q (\mu+\nu)^{N_2} (\mu-\nu)^{N_3} Q_2[N_4 - N_1, p_1(\mu-\nu)] e^{-p_2\mu - p_3\nu} d\mu\, d\nu \tag{39} $$

$$ {}^{Q_3}\mathcal{G}^{N_1,q}_{N_2 N_3 N_4}(p_1, p_2, p_3) = \frac{p_1^{N_1}}{(N_4 - N_1)_{N_1}} \int_0^\infty \int_{-1}^1 (\mu\nu)^q (\mu+\nu)^{N_2} (\mu-\nu)^{N_3} Q_3[N_4 - N_1, p_1(\mu\nu)] e^{-p_2\mu - p_3\nu} d\mu\, d\nu, \tag{40} $$

here, $(a)_n$ is the Pochhammer symbol and $N_1 \ge 0$ , $-\infty < N_2 < \infty$ , $[N_3, N_4] > 0$ , $N_1 \in \mathbb{N}; \{N_2, N_3, N_4\} \in \mathbb{R}$ , respectively.

### 3. Results and Discussions

The derived closed form expressions of two-center two-electron Coulomb and hybrid integrals in terms of auxiliary functions ${}^{P_i,Q_i}\mathcal{G}^{N_1,q}$ based on the use of ellipsoidal coordinates can be evaluated efficiently and accurately for arbitrary values of quantum numbers and orbital parameters via recurrence relations presented in Appendices A, B and using numerical Global-adaptive method with Gauss-Kronrod extension.

In order to investigate the accuracy of two-electron integrals through numerical calculation of ${}^{P_i,Q_i}\mathcal{G}^{N_1,q}$ functions, we use the quadrature rule of sub-domain $r$ in a sequence of $n_r$ point quadrature for approximation to integrals of ${}^{P_i,Q_i}\mathcal{G}^{N_1,q;p_1,p_2,p_3}_{N_2N_3N_4}(\mu,\nu)$, where,

$$
{}^{P_1,Q_1}\mathcal{G}^{N_1,q;p_1,p_2,p_3}_{N_2N_3N_4}(\mu,\nu)=(\mu\nu)^q(\mu+\nu)^{N_2}(\mu-\nu)^{N_3}\left\{\begin{matrix}P_1[N_4-N_1,p_1(\mu+\nu)]\\ Q_1[N_4-N_1,p_1(\mu+\nu)]\end{matrix}\right\}e^{-p_2\mu-p_3\nu} \tag{41}
$$

$$
{}^{P_2,Q_2}\mathcal{G}^{N_1,q;p_1,p_2,p_3}_{N_2N_3N_4}(\mu,\nu)=(\mu\nu)^q(\mu+\nu)^{N_2}(\mu-\nu)^{N_3}\left\{\begin{matrix}P_2[N_4-N_1,p_1(\mu-\nu)]\\ Q_2[N_4-N_1,p_1(\mu-\nu)]\end{matrix}\right\}e^{-p_2\mu-p_3\nu} \tag{42}
$$

$$
{}^{P_3,Q_3}\mathcal{G}^{N_1,q;p_1,p_2,p_3}_{N_2N_3N_4}(\mu,\nu)=(\mu\nu)^q(\mu+\nu)^{N_2}(\mu-\nu)^{N_3}\left\{\begin{matrix}P_3[N_4-N_1,p_1(\mu\nu)]\\ Q_3[N_4-N_1,p_1(\mu\nu)]\end{matrix}\right\}e^{-p_2\mu-p_3\nu} \tag{43}
$$

on intervals $[1,\infty)\times[-1,1]$ determined by,

$$
I=\sum_{s_1=1}^{n_{r1}}\sum_{s_2=1}^{n_{r2}}\omega_{r_1s_1}\omega_{r_2s_2}\,{}^{P_i,Q_i}\mathcal{G}^{N_1,q;p_1,p_2,p_3}_{N_2N_3N_4}(\mu_{r_1s_1},\nu_{r_2s_2}). \tag{44}
$$

Here, $\omega_{rs}$ are the weights , $\{\mu_{rs},\nu_{rs}\}$ are roots and their choice so that $I\approx{}^{P_i,Q_i}\mathcal{G}^{N_1,q}_{N_2N_3N_4}$ define the rule and provide both an integral estimate and an error $(\varepsilon_r)$ estimate as a measure of the integral estimate accuracy (Please see Refs. [20, 55-57] for the details on the implementation of the procedure and information on the roots and singularities).

The algorithm described in Eq.(44), has been incorporated into a computer program written in the Mathematica programming language with the included numerical computation packages. Notice that, the Mathematica program language can handle approximate real numbers with any number of digits and it is suitable only for benchmarking in the view of the calculation times. It gives 25 accurate-decimals in minutes and more than 10 accurate-decimals in seconds. It is designed to provide bechmark integral values.

The results of calculation in atomic units are given in tables 1-7 and figures 1 and 2 for arbitrary values of integer and non-integer quantum numbers with different values of orbital parameters. The benchmark results are obtained by implementing the numerical Global-adaptive method based on the Gauss-Kronrod extension with the maximum recursion and working precision are 35, 25 for tables 1-4 and 5-7, respectively. In these tables, the values entered in the first line are always the benchmark results given with 35 and 25 accurate-decimals for tables 1-4 and 5-7, respectively. Further contents are the results found in the literature or results obtained from alternative methods which are eligible to make comparison.

In table 1 the results obtained for ${}^{P_i,Q_i}\mathcal{G}^{N_1,q}_{N_2 N_3 N_4}$, auxiliary functions where, $N_i \in \mathbb{N}^+$, to be used for STOs are presented. The comparison is made for $\mathcal{G}^{0,q}_{N_2 N_3 N_4}$ functions which are a special case of ${}^{P_i,Q_i}\mathcal{G}^{N_1,q}_{N_2 N_3 N_4}$ for $N_1 = 0$. It can be seen from the table 1 that the strategy applied in this paper for numerical calculation of these functions gives exactly the same results as analytical calculation based on recurrence relations and series representation formulas given in [58-61] which are also special case of recurrence relations given in Appendices A and B and series representation formulas given in Appendix C for $N_i \in \mathbb{N}^+$, respectively. Table 1 also shows that the accuracy of numerical calculations is valid for all cases. In the case of integer values of quantum numbers the series representation formulas can be used in the calculations for small values of parameters. The calculations for large values of parameters should be performed with analytical formulas based on the recurrence relations. It should also be noted that, the numerical calculations for large values of quantum numbers are better suited to the algorithm and faster.

The components in the analytical closed form expressions given for two-electron integrals other than the auxiliary functions are coefficients and no computational problem emerges other than accurate calculation of them. The expression Eq. (88) given in Appendix C, which is the general expression for analytical calculation with an infinite series representation obtained from binomial expansion (Eq. (5)) in the case of non-integer values, when it is considered together with Eqs. (82-87)can not be expected to obtain accurate results from the calculations over NSTOs for all values of parameters. In table 2 the results based on the calculation of ${}^{P_i,Q_i}\mathcal{G}^{N_1,q}_{N_2 N_3 N_4}$ auxiliary functions through numerical Global-adaptive method with Gauss-Kronrod extension where, $N_i \in \mathbb{R}^+$, are presented.

Some values of the two-center Coulomb and hybrid integrals over STOs obtained from Eqs. (33, 34) are presented in the tables 3 and 4 in order to support the reliability of the procedure. As can be seen from these tables the results obtained are in good agreement with values from relations given in [53, 62, 63] and values found in the literature.

In the tables 5, 6 and 7 the results are given for the two-center Coulomb and hybrid integrals over NSTOs, respectively.

In the table 5, convergence of the one-center expansion method is tested. In this table, the upper limit of summation $\left(N_e\right)$ given for one-center expansion method is chosen as 12, 13, 14, 15, 20, 30, respectively. Notice that, the computing time increases with the upper limit of summation. The results are obtained for the highest limit in days on a PC (Intel core i5-3.2Ghz) running the Mathematica platform. It can be seen from this table, using one-center expansion method to the calculation of these integrals gives only a few accurate digits. Its convergence a slow to the numerical values and it requires prohibitive cpu times.

The one-center expansion method is also performed for the two-center Coulomb and hybrid integrals with the given values of quantum numbers and orbital parameters in tables 6 and 7. It is observed that the same number of accurate digits (4-6) are obtained in table 5 for low values of quantum numbers and the number of accurate digits decreases on increasing the values of quantum numbers for the given upper limit of summations in the table 5. Therefore, the results obtained from one-center expansion method are not presented in the table 6 and 7. The benchmark results obtained from numerical Global-adaptive method with Gauss-Kronrod extension are presented in tables 6 and 7 for NSTOs. Comparison is made with Cuba-cuhre numerical integration algorithm on the Mathematica platform. The input parameters are determined as: the accuracy goal, maximum points, precision goal, 50, E+5, E+3, respectively. It can be seen from these tables, the Cuba-cuhre algorithm gives better results than the one-center expansion and it gives up to about 10-12 digit accuracy for low values of orbital parameters.

The convergence properties of series representation formulas of ${}^{P_i,Q_i}\mathcal{G}^{N_1,q}_{N_2N_3N_4}$ auxiliary functions for $N_i \in \mathbb{N}^+$ are examined in the figure 1 and 2 depending on $p_1$ and upper limit of summation $\left(N_s\right)$ , respectively. In the figure 1 the upper limit of summation is chosen as $N_s=100$ . And, in the figure 2 vertical lines show the critical values of $N_s$ which emphasize the ending of oscillation.

The figures 1 and 2 show that, the values of reduced auxiliary functions diverge from accurate numerical values when increasing the parameter $p_1$ and for large values of parameters require increase of upper limits of summations.

Finally, we would like to say that, the formulas and algorithm presented in this paper give highly accurate results for the two-center Coulomb and hybrid integrals and they can be useful in both relativistic and non-relativistic electronic structure calculations for determination of various properties of molecules when HFR, DHFR (Dirac-Hartree-Fock-Roothaan) theories and explicitly correlated approximations are used. Calculation for these integrals is performed without assuming hermiticity, which improves the accuracy. In the future, we plan to perform the method presented in the relativistic electronic structure calculation of diatomic molecules using exponential-type orbitals and present accurate solution for positive- and negative-energy states.

## Appendix A: Recurrence Relations for ${}^{P_i}\mathcal{G}^{N,q}$ and ${}^{Q_i}\mathcal{G}^{N,q}$ Auxiliary Functions

In order to obtain recurrence relations for ${}^{P_i,Q_i}\mathcal{G}^{N,q}$ auxiliary functions the normalized complete gamma functions and the normalized complementary incomplete gamma functions, their upward, downward consecutive neighbors recurrence relations and the first derivatives defined as [64-66],

$$P_i[a,bz_i]=\frac{\gamma(a,bz_i)}{\Gamma[a]};Q_i[a,bz_i]=\frac{\Gamma(a,bz_i)}{\Gamma[a]}, \tag{45}$$

where, $\gamma[a,bz_i]$, $\Gamma[a,bz_i]$ are incomplete gamma functions, with $a,b>0,z_i>0$,

the derivatives,

$$\frac{d}{dz}P_i[a,bz_i]=\frac{b^a}{\Gamma[a]}e^{-bz_i}z_i^{a-1};\ \frac{d}{dz}Q_i[a,bz_i]=-\frac{b^a}{\Gamma[a]}e^{-bz_i}z_i^{a-1}, \tag{46}$$

the upward recurrence relations,

$$P_i[a,bz_i]=P_i[a+1,bz_i]+\frac{b^a}{\Gamma[a+1]}e^{-bz_i}z_i^a;\ Q_i[a,bz_i]=Q_i[a+1,bz_i]-\frac{b^a}{\Gamma[a+1]}e^{-bz_i}z_i^a, \tag{47}$$

and the downward recurrence relations,

$$P_i[a,bz_i]=P_i[a-1,bz_i]-\frac{b^{a-1}}{\Gamma[a]}e^{-bz_i}(z_i)^{a-1};Q_i[a,bz_i]=Q_i[a-1,bz_i]+\frac{b^{a-1}}{\Gamma[a]}e^{-bz_i}(z_i)^{a-1}. \tag{48}$$

Notice that, $P_i$ and $Q_i$ satisfy the identity $P_i+Q_i=1$.

Taking into consideration these properties of incomplete gamma functions and applying integration by parts over $\nu$, the recurrence relations for ${}^{P_i}\mathcal{G}^{N_1,q}$ and ${}^{Q_i}\mathcal{G}^{N_1,q}$ auxiliary functions can be obtained as follows:

Firstly, using,

$$(\mu\nu)=\frac{1}{4}\left\{(\mu+\nu)^2-(\mu-\nu)^2\right\}, \tag{49}$$

$$(\mu\nu)^2=\frac{1}{16}\left\{(\mu+\nu)^4+2(\mu+\nu)^2(\mu-\nu)^2+(\mu-\nu)^4\right\}, \tag{50}$$

the relations for auxiliary functions can be reduced to ${}^{P_i,Q_i}\mathcal{G}^{N_1,0}$,

$$ {}^{P_i,Q_i}\mathcal{G}^{N_1,q}_{N_2N_3N_4}(p_1,p_2,p_3)=\frac{1}{4}\left\{{}^{P_i,Q_i}\mathcal{G}^{N_1,q-1}_{N_2+2N_3N_4}(p_1,p_2,p_3)-{}^{P_i,Q_i}\mathcal{G}^{N_1,q-1}_{N_2N_3+2N_4}(p_1,p_2,p_3)\right\};\ q\geq 1, \tag{51}$$

$$ \begin{aligned}{}^{P_i,Q_i}\mathcal{G}^{N_1,q}_{N_2N_3N_4}(p_1,p_2,p_3)=\frac{1}{16}\Big\{&{}^{P_i,Q_i}\mathcal{G}^{N_1,q-2}_{N_2+4N_3N_4}(p_1,p_2,p_3)-2\,{}^{P_i,Q_i}\mathcal{G}^{N_1,q-2}_{N_2+2N_3+2N_4}(p_1,p_2,p_3)+\\ &{}^{P_i,Q_i}\mathcal{G}^{N_1,q-2}_{N_2N_3+4N_4}(p_1,p_2,p_3)\Big\};\ q\geq 2,\end{aligned} \tag{52}$$

and, according to recurrence relations of gamma functions, the upward and the downward relations for ${}^{P_i,Q_i}\mathcal{G}^{N_1,q}$, functions over $N_1, N_4$ given as, the upward relations,

$$ \begin{aligned}{}^{P_1,Q_1}\mathcal{G}^{N_1,0}_{N_2N_3N_4}(p_1,p_2,p_3)=&\left(\frac{N_4}{N_4-N_1}\right)\Big\{{}^{P_1,Q_1}\mathcal{G}^{N_1,0}_{N_2N_3N_4+1}(p_1,p_2,p_3)\pm\\ &\pm\frac{p_1^{N_4-N_1}}{\Gamma[N_4-N_1+1]}\mathcal{G}^{N_1,0}_{N_2+N_4-N_1N_3N_4+1}(p_1,p_2+p_1,p_3+p_1)\Big\},\end{aligned} \tag{53}$$

$$ \begin{aligned}{}^{P_2,Q_2}\mathcal{G}^{N_1,0}_{N_2N_3N_4}(p_1,p_2,p_3)=&\left(\frac{N_4}{N_4-N_1}\right)\Big\{{}^{P_2,Q_2}\mathcal{G}^{N_1,0}_{N_2N_3N_4+1}(p_1,p_2,p_3)\pm\\ &\pm\frac{p_1^{N_4-N_1}}{\Gamma[N_4-N_1+1]}\mathcal{G}^{N_1,0}_{N_2N_3+N_4-N_1N_4+1}(p_1,p_2+p_1,p_3-p_1)\Big\},\end{aligned} \tag{54}$$

$$ \begin{aligned}{}^{P_3,Q_3}\mathcal{G}^{N_1,0}_{N_2N_3N_4}(p_1,p_2,p_3)=&\left(\frac{N_4}{N_4-N_1}\right)\Big\{{}^{P_3,Q_3}\mathcal{G}^{N_1,0}_{N_2N_3N_4+1}(p_1,p_2,p_3)\pm\\ &\pm\frac{p_1^{N_4-N_1}}{\Gamma[N_4-N_1+1]}\mathcal{G}^{N_1,N_4-N_1}_{N_2N_3N_4+1}(p_1,p_2+p_1\nu,p_3)\Big\},\end{aligned} \tag{55}$$

and the downward relations,

$$ {}^{P_1,Q_1}\mathcal{G}^{N_1,0}_{N_2N_3N_4}(p_1,p_2,p_3)=\left(\frac{N_4-N_1-1}{N_4-1}\right)\left\{{}^{P_1,Q_1}\mathcal{G}^{N_1,0}_{N_2N_3N_4-1}(p_1,p_2,p_3)\mp \right. $$
$$ \left. \mp\frac{p_1^{N_4-N_1-1}}{\Gamma[N_4-N_1]}\mathcal{G}^{N_1,0}_{N_2+N_4-N_1-1N_3N_4-1}(p_1,p_2+p_1,p_3+p_1)\right\}, \tag{56} $$

$$ {}^{P_2,Q_2}\mathcal{G}^{N_1,0}_{N_2N_3N_4}(p_1,p_2,p_3)=\left(\frac{N_4-N_1-1}{N_4-1}\right)\left\{{}^{P_2,Q_2}\mathcal{G}^{N_1,0}_{N_2N_3N_4-1}(p_1,p_2,p_3)\mp \right. $$
$$ \left. \mp\frac{p_1^{N_4-N_1-1}}{\Gamma[N_4-N_1]}\mathcal{G}^{N_1,0}_{N_2N_3+N_4-N_1-1N_4-1}(p_1,p_2+p_1,p_3-p_1)\right\}, \tag{57} $$

$$ {}^{P_3,Q_3}\mathcal{G}^{N_1,0}_{N_2N_3N_4}(p_1,p_2,p_3)=\left(\frac{N_4-N_1-1}{N_4-1}\right)\left\{{}^{P_3,Q_3}\mathcal{G}^{N_1,0}_{N_2N_3N_4-1}(p_1,p_2,p_3)\mp \right. $$
$$ \left. \mp\frac{p_1^{N_4-N_1-1}}{\Gamma[N_4-N_1]}\mathcal{G}^{N_1,N_4-N_1-1}_{N_2N_3N_4-1}(p_1,p_2+p_1\nu,p_3)\right\}, \tag{58} $$

where, $\mathcal{G}^{N_1,q}$ are ${}^{P_i,Q_i}\mathcal{G}^{N_1,q}$ -like reduced-auxiliary functions, which are independent from gamma functions,

$$ \mathcal{G}^{N_1,q}_{N_2N_3N_4}(p_1,p_2,p_3)=\frac{p_1^{N_1}}{(N_4-N_1)_{N_1}}\int_0^\infty\int_{-1}^{1}(\mu\nu)^q(\mu+\nu)^{N_2}(\mu-\nu)^{N_3}e^{-p_2\mu-p_3\nu}d\mu\, d\nu. \tag{59} $$

Notice that, the $\mathcal{G}^{N_1,q}$ auxiliary functions are different from auxiliary function given in [58, 60] by the product term on the left-hand side.

Considering the derivatives of gamma functions and making use of integration by parts, the recurrence relations over $N_2, N_3$ can be obtained as,

$$ {}^{P_1,Q_1}\mathcal{G}^{N_1,0}_{N_2N_3N_4}(p_1,p_2,p_3)=\frac{1}{p_2}\left\{e^{-p_2}\ {}^{P_1,Q_1}\mathcal{J}^{N_1,0}_{N_2N_3N_4}(p_1,p_3)+N_2\ {}^{P_1,Q_1}\mathcal{G}^{N_1,0}_{N_2-1N_3N_4}(p_1,p_2,p_3)+ \right. $$
$$ \left. +N_3\ {}^{P_1,Q_1}\mathcal{G}^{N_1,0}_{N_2N_3-1N_4}(p_1,p_2,p_3)\pm\frac{p_1^{N_4-N_1}}{\Gamma[N_4-N_1]}\mathcal{G}^{N_1,0}_{N_2+N_4-N_1-1N_3N_4}(p_1,p_2+p_1,p_3+p_1)\right\}, \tag{60} $$

$$ {}^{P_2,Q_2}\mathcal{G}^{N_1,0}_{N_2N_3N_4}(p_1,p_2,p_3)=\frac{1}{p_2}\Big\{e^{-p_2}\ {}^{P_2,Q_2}\mathcal{J}^{N_1,0}_{N_2N_3N_4}(p_1,p_3)+N_2\ {}^{P_2,Q_2}\mathcal{G}^{N_1,0}_{N_2-1N_3N_4}(p_1,p_2,p_3)+ $$
$$ +N_3\ {}^{P_2,Q_2}\mathcal{G}^{N_1,0}_{N_2N_3-1N_4}(p_1,p_2,p_3)\pm\frac{p_1^{N_4-N_1}}{\Gamma[N_4-N_1]}\mathcal{G}^{N_1,0}_{N_2N_3+N_4-N_1-1N_4}(p_1,p_2+p_1,p_3-p_1)\Big\}, \tag{61} $$

$$ {}^{P_3,Q_3}\mathcal{G}^{N_1,0}_{N_2N_3N_4}(p_1,p_2,p_3)=\frac{1}{p_2}\Big\{e^{-p_2}\ {}^{P_3,Q_3}\mathcal{J}^{N_1,0}_{N_2N_3N_4}(p_1,p_3)+N_2\ {}^{P_3,Q_3}\mathcal{G}^{N_1,0}_{N_2-1N_3N_4}(p_1,p_2,p_3)+ $$
$$ +N_3\ {}^{P_3,Q_3}\mathcal{G}^{N_1,0}_{N_2N_3-1N_4}(p_1,p_2,p_3)\pm\frac{p_1^{N_4-N_1}}{\Gamma[N_4-N_1]}\mathcal{G}^{N_1,N_4-N_1-1}_{N_2N_3N_4}(p_2+p_1\nu,p_3)\Big\}. \tag{62} $$

The one-variable ${}^{P_i,Q_i}\mathcal{J}^{N_1,q}$ auxiliary functions occurring in Eqs. (60-62) are determined as follows,

$$ {}^{P_1Q_1}\mathcal{J}^{N_1,q}_{N_2N_3N_4}(p_1,p_3)=\frac{p_1^{N_1}}{(N_4-N_1)_{N_1}}\int_{-1}^{1}\nu^q(1+\nu)^{N_2}(1-\nu)^{N_3}\begin{Bmatrix}P_1[N_4-N_1,p_1(1+\nu)]\\Q_1[N_4-N_1,p_1(1+\nu)]\end{Bmatrix}e^{-p_3\nu}d\nu, \tag{63} $$

$$ {}^{P_2Q_2}\mathcal{J}^{N_1,q}_{N_2N_3N_4}(p_1,p_3)=\frac{p_1^{N_1}}{(N_4-N_1)_{N_1}}\int_{-1}^{1}\nu^q(1+\nu)^{N_2}(1-\nu)^{N_3}\begin{Bmatrix}P_1[N_4-N_1,p_1(1-\nu)]\\Q_1[N_4-N_1,p_1(1-\nu)]\end{Bmatrix}e^{-p_3\nu}d\nu, \tag{64} $$

$$ {}^{P_3Q_3}\mathcal{J}^{N_1,q}_{N_2N_3N_4}(p_1,p_3)=\frac{p_1^{N_1}}{(N_4-N_1)_{N_1}}\int_{-1}^{1}\nu^q(1+\nu)^{N_2}(1-\nu)^{N_3}\begin{Bmatrix}P_1[N_4-N_1,p_1\nu]\\Q_1[N_4-N_1,p_1\nu]\end{Bmatrix}e^{-p_3\nu}d\nu. \tag{65} $$

It should be noted that, The recurrence relations of ${}^{P_i,Q_i}\mathcal{J}^{N_1,q}$ auxiliary functions can be obtained following same method used for ${}^{P_i,Q_i}\mathcal{G}^{N_1,q}$ functions.

### Appendix B: Recurrence Relations for ${}^{P_i}\mathcal{J}^{N,q}$ and ${}^{Q_i}\mathcal{J}^{N,q}$ Auxiliary Functions

Starting from the recurrence relations of gamma functions again, the upward and downward relationships for ${}^{P_i,Q_i}\mathcal{J}^{N_1,q}$ functions over $N_1, N_4$ can be obtained as,

$$ {}^{P_i,Q_i}\mathcal{J}^{N_1,q}_{N_2N_3N_4}(p_1,p_3)=\frac{1}{4}\Big\{{}^{P_i,Q_i}\mathcal{J}^{N_1,q-1}_{N_2+2N_3N_4}(p_1,p_3)-{}^{P_i,Q_i}\mathcal{J}^{N_1,q-1}_{N_2N_3+2N_4}(p_1,p_3)\Big\};\ q\geq1. \tag{66} $$

$$ {}^{P_i,Q_i}\mathcal{J}^{N_1,q}_{N_2N_3N_4}(p_1,p_2,p_3)=\frac{1}{16}\Big\{{}^{P_i,Q_i}\mathcal{J}^{N_1,q-2}_{N_2+4N_3N_4}(p_1,p_2,p_3)-2\ {}^{P_i,Q_i}\mathcal{J}^{N_1,q-2}_{N_2+2N_3+2N_4}(p_1,p_2,p_3)+ $$
$$ {}^{P_i,Q_i}\mathcal{J}^{N_1,q-2}_{N_2N_3+4N_4}(p_1,p_2,p_3)\Big\};\ q\geq2, \tag{67} $$

the upward relationships,

$$
\begin{aligned}
{}^{P_1,Q_1}\mathcal{J}^{N_1,0}_{N_2N_3N_4}(p_1,p_3)=&\left(\frac{N_4}{N_4-N_1}\right)\left\{{}^{P_1,Q_1}\mathcal{J}^{N_1,0}_{N_2N_3N_4+1}(p_1,p_3)\pm\right.\\
&\left.\pm\frac{p_1^{N_4-N_1}}{\Gamma[N_4-N_1+1]}e^{-p_1}\mathcal{J}^{N_1,0}_{N_2+N_4-N_1N_3N_4+1}(p_3+p_1)\right\},
\end{aligned}
\tag{68}
$$

$$
\begin{aligned}
{}^{P_2,Q_2}\mathcal{J}^{N_1,0}_{N_2N_3N_4}(p_1,p_3)=&\left(\frac{N_4}{N_4-N_1}\right)\left\{{}^{P_2,Q_2}\mathcal{J}^{N_1,0}_{N_2N_3N_4+1}(p_1,p_3)\pm\right.\\
&\left.\pm\frac{p_1^{N_4-N_1}}{\Gamma[N_4-N_1+1]}e^{-p_1}\mathcal{J}^{N_1,0}_{N_2N_3+N_4-N_1N_4+1}(p_3-p_1)\right\},
\end{aligned}
\tag{69}
$$

$$
\begin{aligned}
{}^{P_3,Q_3}\mathcal{J}^{N_1,0}_{N_2N_3N_4}(p_1,p_3)=&\left(\frac{N_4}{N_4-N_1}\right)\left\{{}^{P_3,Q_3}\mathcal{J}^{N_1,0}_{N_2N_3N_4+1}(p_1,p_3)\pm\right.\\
&\left.\pm\frac{p_1^{N_4-N_1}}{\Gamma[N_4-N_1+1]}\mathcal{J}^{N_1,N_4-N_1}_{N_2N_3N_4+1}(p_1+p_3)\right\},
\end{aligned}
\tag{70}
$$

and the downward relationships,

$$
\begin{aligned}
{}^{P_1,Q_1}\mathcal{J}^{N_1,0}_{N_2N_3N_4}(p_1,p_3)=&\left(\frac{N_4-N_1-1}{N_4-1}\right)\left\{{}^{P_1,Q_1}\mathcal{J}^{N_1,0}_{N_2N_3N_4-1}(p_1,p_3)\mp\right.\\
&\left.\mp\frac{p_1^{N_4-N_1-1}}{\Gamma[N_4-N_1]}e^{-p_1}\mathcal{J}^{N_1,0}_{N_2+N_4-N_1-1N_3N_4-1}(p_3+p_1)\right\},
\end{aligned}
\tag{71}
$$

$$
\begin{aligned}
{}^{P_2,Q_2}\mathcal{J}^{N_1,0}_{N_2N_3N_4}(p_1,p_3)=&\left(\frac{N_4-N_1-1}{N_4-1}\right)\left\{{}^{P_2,Q_2}\mathcal{J}^{N_1,0}_{N_2N_3N_4-1}(p_1,p_3)\mp\right.\\
&\left.\mp\frac{p_1^{N_4-N_1-1}}{\Gamma[N_4-N_1]}e^{-p_1}\mathcal{J}^{N_1,0}_{N_2N_3+N_4-N_1-1\,N_4-1}(p_3-p_1)\right\},
\end{aligned}
\tag{72}
$$

$$
\begin{aligned}
{}^{P_3,Q_3}\mathcal{J}^{N_1,0}_{N_2N_3N_4}(p_1,p_3)=&\left(\frac{N_4-N_1-1}{N_4-1}\right)\left\{{}^{P_3,Q_3}\mathcal{J}^{N_1,0}_{N_2N_3N_4-1}(p_1,p_3)\mp\right.\\
&\left.\mp\frac{p_1^{N_4-N_1-1}}{\Gamma[N_4-N_1]}\mathcal{J}^{N_1,N_4-N_1-1}_{N_2N_3N_4-1}(p_1+p_3)\right\},
\end{aligned}
\tag{73}
$$

where, $\mathcal{J}^{N_1,q}$ are ${}^{P_i,Q_i}\mathcal{J}^{N_1,q}$ -like reduced-auxiliary functions, which are also independent from gamma functions,

$$
\mathcal{J}^{N_1,q}_{N_2N_3N_4}(p_1,p_3)=\frac{p_1^{N_1}}{(N_4-N_1)_{N_1}}\int_{-1}^{1}\nu^q(1+\nu)^{N_2}(1-\nu)^{N_3}e^{-p_3\nu}d\nu. \tag{74}
$$

In the same manner by the use of the derivatives of gamma functions and using integration by parts, the relationships over $N_2, N_3$ can be obtained as,

$$
\begin{aligned}
{}^{P_1,Q_1}\mathcal{J}^{N_1,0}_{N_2N_3N_4}(p_1,p_3) &= \frac{1}{p_3}\Big\{N_2\,{}^{P_1,Q_1}\mathcal{J}^{N_1,0}_{N_2-1N_3N_4}(p_1,p_3) - N_3\,{}^{P_1,Q_1}\mathcal{J}^{N_1,0}_{N_2N_3-1N_4}(p_1,p_3) \pm \\
&\frac{p_1^{N_4-N_1}}{\Gamma[N_4-N_1]} e^{-p_1}\mathcal{J}^{N_1,0}_{N_2+N_4-N_1-1N_3N_4}(p_1,p_3+p_1) - {}^{P_1,Q_1}J^{N_1}_{N_2N_3N_4}(p_1,p_3),
\end{aligned}
\tag{75}
$$

$$
\begin{aligned}
{}^{P_2,Q_2}\mathcal{J}^{N_1,0}_{N_2N_3N_4}(p_1,p_3) &= \frac{1}{p_3}\Big\{N_2\,{}^{P_2,Q_2}\mathcal{J}^{N_1,0}_{N_2-1N_3N_4}(p_1,p_3) - N_3\,{}^{P_2,Q_2}\mathcal{J}^{N_1,0}_{N_2N_3-1N_4}(p_1,p_3) \pm \\
&\frac{p_1^{N_4-N_1}}{\Gamma[N_4-N_1]} e^{-p_1}\mathcal{J}^{N_1,0}_{N_2N_3+N_4-N_1-1N_4}(p_1,p_3-p_1) - {}^{P_2,Q_2}J^{N_1}_{N_2N_3N_4}(p_1,p_3),
\end{aligned}
\tag{76}
$$

$$
\begin{aligned}
{}^{P_3,Q_3}\mathcal{J}^{N_1,0}_{N_2N_3N_4}(p_1,p_3) &= \frac{1}{p_3}\Big\{N_2\,{}^{P_3,Q_3}\mathcal{J}^{N_1,0}_{N_2-1N_3N_4}(p_1,p_3) - N_3\,{}^{P_3,Q_3}\mathcal{J}^{N_1,0}_{N_2N_3-1N_4}(p_1,p_3) \pm \\
&\frac{p_1^{N_4-N_1}}{\Gamma[N_4-N_1]} \mathcal{J}^{N_1,N_4-N_1-1}_{N_2N_3N_4}(p_1,p_3+p_1) - {}^{P_3,Q_3}J^{N_1}_{N_2N_3N_4}(p_1,p_3).
\end{aligned}
\tag{77}
$$

Finally, the resulting ${}^{P_i,Q_i}J^{N_1}$ coefficients are defined as follows,

$$
{}^{P_1,Q_1}J^{N_1,0}_{N_2N_3N_4}(p_1,p_3) = \frac{p_1^{N_1}}{(N_4-N_1)_{N_1}}\left[2^{N_2}\delta_{N_30}\begin{Bmatrix}P_1[N_4-N_1,2\mathrm{p}_1]\\ Q_1[N_4-N_1,2\mathrm{p}_1]\end{Bmatrix} - 2^{N_3}\delta_{N_20}\begin{Bmatrix}0\\1\end{Bmatrix}\right]e^{-p_3}
\tag{78}
$$

$$
{}^{P_2,Q_2}J^{N_1,0}_{N_2N_3N_4}(p_1,p_3) = \frac{p_1^{N_1}}{(N_4-N_1)_{N_1}}\left[2^{N_2}\delta_{N_30}\begin{Bmatrix}0\\1\end{Bmatrix} - 2^{N_3}\delta_{N_20}\begin{Bmatrix}P_1[N_4-N_1,2\mathrm{p}_1]\\ Q_1[N_4-N_1,2\mathrm{p}_1]\end{Bmatrix}\right]e^{-p_3}
\tag{79}
$$

$$
\begin{aligned}
{}^{P_3,Q_3}J^{N_1,0}_{N_2N_3N_4}(p_1,p_3) &= \frac{p_1^{N_1}}{(N_4-N_1)_{N_1}}\Bigg[2^{N_2}\delta_{N_30}\begin{Bmatrix}P_1[N_4-N_1,p_1]\\ Q_1[N_4-N_1,p_1]\end{Bmatrix} - \\
&-2^{N_3}\delta_{N_20}\begin{Bmatrix}P_1[N_4-N_1,-p_1]\\ Q_1[N_4-N_1,-p_1]\end{Bmatrix}\Bigg]e^{-p_3}.
\end{aligned}
\tag{80}
$$

The integration by parts can also be performed over $\mu$, the expression found for ${}^{P_i,Q_i}\mathcal{G}^{N_1,q}$, ${}^{P_i,Q_i}\mathcal{J}^{N_1,q}$ auxiliary functions can be combined to obtain new sets of relations, which may be better suited amount of computation time. The ${}^{P_i,Q_i}\mathcal{G}^{N_1,q}$ functions as they are differ from auxiliary functions given in [58, 61] by product term they can be used to perform the calculations over STOs with their presented recurrence relations in this paper.

The incomplete gamma functions $\left(\Gamma\left[a, bz_i\right]\right)$ in the region $0 \leq a < 1$ are unstable. Generation of the incomplete gamma functions by means of recurrence relations for $0 \leq a < 1$ in an efficient way and computing the gamma functions without erroneous last digits is still being studied in the literature [46, 64, 67-69]. This poses a major problem in the evaluation of auxiliary functions based on the recurrence relation for NSTOs as the terms with lowest values of quantum numbers and orbital parameters should describe with infinite series representation [70]. In the present paper the given recurrence relations are used for lowering the indices in the case of large quantum numbers and numerical Global-adaptive method is performed on the auxiliary functions which have lowered quantum numbers for NSTOs. Notice that, the computational procedure applied here for these auxiliary functions gives more than 15 accurate-decimals for Coulomb and hybrid integrals in seconds using Mathematica programming language.

### Appendix C: Analytical Evaluation ${}^{P_i}\mathcal{G}^{N,q}$ and ${}^{Q_i}\mathcal{G}^{N,q}$ Auxiliary Functions

In order to express the functions ${}^{P_i}\mathcal{G}^{N,q}$ and ${}^{Q_i}\mathcal{G}^{N,q}$ in terms of Mulliken auxiliary functions, it is obtained from following the series representation relation of complete Gamma functions [66],

$$P\left[a, x\right] = e^{-x} \sum_{s=0}^{\infty} \frac{x^{a+s}}{\Gamma\left[a+s+1\right]}, \tag{81}$$

and analytical expressions given in [19]. Thus, the analytical relations for these functions based on the infinite series representation expressions can be obtained as follows, for ${}^{P_1,Q_1}\mathcal{G}^{N,q}$,

$${}^{P_1}\mathcal{G}^{N_1,q}_{N_2N_3N_4} = \frac{p_1^{N_1}}{\left(N_4 - N_1\right)_{N_1}} \sum_{s=0}^{\infty} \frac{p_1^s}{\Gamma\left[N_4 - N_1 + s + 1\right]} \mathcal{G}^{0,q}_{N_2+sN_30}\left(0, p_2 + p_1, p_3 + p_1\right), \tag{82}$$

$${}^{Q_1}\mathcal{G}^{N_1,q}_{N_2N_3N_4} = \frac{p_1^{N_1}}{\left(N_4 - N_1\right)_{N_1}} \left\{ \mathcal{G}^{0,q}_{N_2N_30}\left(0, p_2, p_3\right) - \sum_{s=0}^{\infty} \frac{p_1^s}{\Gamma\left[N_4 - N_1 + s + 1\right]} \mathcal{G}^{0,q}_{N_2+sN_30}\left(0, p_2 + p_1, p_3 + p_1\right) \right\}, \tag{83}$$

for ${}^{P_2,Q_2}\mathcal{G}^{N,q}$,

$${}^{P_2}\mathcal{G}^{N_1,q}_{N_2N_3N_4} = \frac{p_1^{N_1}}{\left(N_4 - N_1\right)_{N_1}} \sum_{s=0}^{\infty} \frac{p_1^s}{\Gamma\left[N_4 - N_1 + s + 1\right]} \mathcal{G}^{0,q}_{N_2N_3+s0}\left(0, p_2 + p_1, p_3 - p_1\right), \tag{84}$$

$$ {}^{Q_2}\mathcal{G}^{N_1,q}_{N_2N_3N_4}=\frac{p_1^{N_1}}{(N_4-N_1)_{N_1}}\left\{\mathcal{G}^{0,q}_{N_2N_30}(0,p_2,p_3)-\sum_{s=0}^{\infty}\frac{p_1^s}{\Gamma[N_4-N_1+s+1]}\mathcal{G}^{0,q}_{N_2N_3+s\,0}(0,p_2+p_1,p_3-p_1)\right\}, \quad (85) $$

and for ${}^{P_3,Q_3}\mathcal{G}^{N,q}$,

$$ {}^{P_3}\mathcal{G}^{N_1,q}_{N_2N_3N_4}=\frac{p_1^{N_1}}{(N_4-N_1)_{N_1}}\sum_{s=0}^{\infty}\frac{p_1^s}{\Gamma[N_4-N_1+s+1]}\mathcal{G}^{0,q+s}_{N_2N_30}(0,p_2+p_1\nu,p_3), \quad (86) $$

$$ {}^{Q_1}\mathcal{G}^{N_1,q}_{N_2N_3N_4}=\frac{p_1^{N_1}}{(N_4-N_1)_{N_1}}\left\{\mathcal{G}^{0,q}_{N_2N_30}(0,p_2,p_3)-\sum_{s=0}^{\infty}\frac{p_1^s}{\Gamma[N_4-N_1+s+1]}\mathcal{G}^{0,q+s}_{N_2N_30}(0,p_2\nu+p_1,p_3)\right\}, \quad (87) $$

with,

$$ \mathcal{G}^{0,q}_{N_2N_30}(0,p_2,p_3)=\lim_{s\to\infty}\sum_{k=0}^{N_2+s}F_k^s(N_2,N_3)\,A_{N_2+N_3+q-k}(p_2)B_{q+k}(p_3), \quad (88) $$

where,

$$ A_m(p_2)=\int_1^{\infty}\mu^m e^{-p_2\mu}\,d\mu, \quad (89) $$

$$ B_m(p_3)=\int_{-1}^{1}\nu^m e^{-p_3}\,d\nu, \quad (90) $$

are the Mulliken auxiliary functions and $F_m^s(N_2,N_3)$ are generalized binomial coefficients,

$$ F_m^s(N_2,N_3)=\sum_{\sigma=0}^{s}(-1)^{\sigma}F_{m-\sigma}(N_2)F_{\sigma}(N_3). \quad (91) $$

**Acknowledgment.**

A. Bağcı would like to acknowledge funding for a postdoctoral research fellowship from innov@pole: the Auvergne Region and FEDER.

**Table 1.** The comparative values of auxiliary functions ${}^{P_1}\mathcal{G}^{0,q}_{N_2 N_3 N_4}$ for $N_i \in \mathbb{N}^+$

| $N_2$ | $N_3$ | $N_4$ | $q$ | $p_1$ | $p_2$ | $p_3$ | $N_s$ | Results |
|---|---|---|---|---|---|---|---|---|
| 1 | 1 | 3 | 1 | 2 | 12 | 1.5 | - | -1.15343 41695 52208 62207 84929 64825 58325 E-07 |
| 1 | 3 | 1 | 1 | 2 | 12 | 1.5 | - | -7.84235 20645 07762 77321 86752 70231 02223 E-06 |
| 2 | 3 | 1 | 1 | 2 | 12 | 1.5 | - | -5.84021 41543 38575 37468 78357 87991 82838 E-05 |
| 2 | 1 | 3 | 3 | 15 | 20 | 12 | - | -3.98124 60701 87560 42411 25094 99407 21556 E-05 |
| 4 | 3 | 5 | 4 | 5 | 4 | 1 | - | 1.013266 52428 99228 68829 33708 47926 62535 E-01 |
| 4 | 5 | 8 | 4 | 0.8 | 20 | 1.4 | - | 4.60128 84384 24297 02627 05703 72641 48936 E-17 |
| 6 | 5 | 6 | 6 | 1.8 | 10 | 1 | 90 | **3.89450 66837 96534 91806 96567 64124 65077** E-06[a, b, c]<br>**3.894**69 48598 4450 E-06[d] |
| 6 | 5 | 7 | 6 | 1.8 | 10 | 1 | - | 2.85306 66419 35908 70824 66940 03248 55651 E-7 |
| 4 | 7 | 4 | 9 | 15 | 20 | 12 | 500 | **-1.27394 18984 52996 23835 63731 40103 73040** E-01[a, b, c]<br>**-1.27394 18984** 5034 E-01[d] |
| 3 | 5 | 3 | 4 | 5 | 4 | 1 | - | **1.43255 70363 69554 53413 90257 62737 65448** E-00[a, b] |
| 6 | 5 | 6 | 4 | 0.8 | 20 | 1.4 | 50 | **7.33615 66598 73843 17340 87048 30849 34519** E-13[a, b, c] |
| 6 | 4 | 6 | 4 | 0.8 | 20 | 14 | 50 | **9.31873 24120 75432 18797 86983 49531 60558** E-08[a, b, c]<br>**9.31873 24120 7543** E-08[d]<br>**9.31873 24120 75432 18797 86983 4953** E-08[e] |
| 8 | 10 | 8 | 8 | 0.07 | 6 | 0.6 | 50 | **4.32621 78828 52319 83727 88650 07840 04716** E-13[a, b, c]<br>**4.32621 78828 523**2 E-13[d]<br>**4.32621 78828 52319 83727 884**29 3831 E-13[e] |
| 10 | 10 | 10 | 10 | 0.03 | 0.8 | 0.08 | 50 | **4.30438 16369 16630 84455 32073 83756 19706** E-03[a, b, c]<br>**4.30438 16369 1663** E-03[d]<br>**4.30438 16369 16630 84455 32073 8376** E-03[e] |
| 17 | 12 | 12 | 16 | 3 | 22 | 15.4 | 50 | **9.69765 04785 20472 28601 82607 01373 18328** E-09[a, c]<br>**9.69765 04785 20472 28601 82607 0137** E-07[e] |
| 20 | 25 | 20 | 25 | 0.6 | 15 | 1.05 | 50 | **-1.38252 33430 90551 19101 89740 33091 84838** E-17[a, c]<br>**-1.38252 33430 90551 19101 89740 3309** E-17[e] |
| 20 | 25 | 20 | 25 | 60 | 15 | 1.05 | - | **-1.3097 30696 86664 04941 60077 87137 77693** E+15[a] |
| 40 | 65 | 40 | 30 | 600 | 15 | 0.15 | - | **6.17374 75959 36901 75370 43063 63073 13455** E+70[a] |
| 65 | 76 | 65 | 66 | 1.2 | 24 | 9.6 | 80 | **4.41492 02430 13616 87153 47681 60265 57629** E-33[a, c] |
| 65 | 76 | 65 | 66 | 120 | 24 | 9.6 | - | **7.71744 47789 00375 89175 90437 82707 02772** E-56[a] |
| 86 | 86 | 86 | 86 | 0.9 | 32 | 22.4 | 60 | **2.38289 43664 90418 64191 74601 04584 83540** E-71[a, c]<br>**2.38289 43664 90418 64191 74601 0458** E-71[e] |
| 86 | 86 | 86 | 86 | 90 | 32 | 22.4 | 60 | **3.04609 07864 62693 92828 68038 48138 07134** E+56[a] |
| 100 | 100 | 100 | 100 | 0.2 | 12 | 9.6 | 60 | **3.03896 54964 16960 07216 15049 58398 40396** E-67[a, c]<br>**3.03896 54964 16960 07**999 546 E-67[e] |
| 100 | 100 | 100 | 100 | 2 | 12 | 9.6 | 250 | **5.42087 65177 37610 22365 38502 04626 00693** E+22[a, c] |

[a]Global-adaptive method with Gauss-Kronrod extension

[b]Recurrence relations given in Appendices A, B with Refs.[58, 60] for $N_i \in \mathbb{N}^+$

[c]Series representation (Eq. (82)) with Ref. [61] for $N_i \in \mathbb{N}^+$

[d]Ref. [59]

[e]Ref. [61]

**Table 2.** The comparative values of auxiliary functions ${}^{P_1}\mathcal{G}^{0,q}_{N_2 N_3 N_4}$ for $N_i \in \mathbb{R}^+$

| $N_2$ | $N_3$ | $N_4$ | $q$ | $p_1$ | $p_2$ | $p_3$ | Results |
|---|---|---|---|---|---|---|---|
| 2.3 | 1.3 | 3.3 | 3 | 15 | 20 | 12 | -7.20013 47001 36267 29748 91419 81660 53601 E-05 |
| 3.6 | 2.6 | 5.2 | 4 | 5 | 4 | 1 | 4.58185 80831 03141 07643 56804 34806 01483 E-02 |
| 4.4 | 4.5 | 7.5 | 4 | 0.8 | 20 | 1.4 | 2.45970 38207 36991 51320 48793 29529 96287 E-16 |
| 6.1 | 5.2 | 6.1 | 6 | 1.8 | 10 | 1 | 3.97200 64539 26573 03355 69307 66129 03949 E-16 |
| 6.1 | 5.2 | 7.2 | 6 | 1.8 | 10 | 1 | 2.26354 74052 81424 11025 57714 62836 32456 E-07 |
| 4.3 | 7.5 | 4.3 | 9 | 15 | 20 | 12 | -2.53202 24148 68428 67810 74097 64810 77155 E-01 |
| 3.6 | 5.4 | 3.6 | 4 | 5 | 4 | 1 | 2.09487 78789 24276 09769 30480 69359 86104 E-00 |
| 6.4 | 5.5 | 6.4 | 4 | 0.8 | 20 | 1.4 | 4.38868 09534 05029 05196 66580 91599 69140 E-13 |
| 6.8 | 6.1 | 6.8 | 4 | 0.8 | 20 | 1.4 | 3.60052 93840 38476 02238 16430 79107 73234 E-08 |
| 8.5 | 10.5 | 8.5 | 8 | 0.07 | 6 | 0.6 | 7.32260 34423 08107 83761 50586 43132 18776 E-14 |
| 8.5 | 10.5 | 8.2 | 8 | 0.07 | 6 | 0.6 | 2.70098 66718 68939 17249 81782 73504 00438 E-33 |
| 10.3 | 10.3 | 10.3 | 10 | 0.03 | 0.8 | 0.08 | 1.94223 81401 94429 27706 51922 22895 90000 E-03 |
| 20.5 | 25.4 | 20.5 | 25 | 0.06 | 15 | 10.5 | -3.48986 70793 88706 99022 28310 61846 41734 E-38 |
| 86.3 | 86.3 | 86.3 | 86 | 0.9 | 32 | 22.4 | 1.02533 51847 51593 55071 76532 03711 21500 E-71 |
| 86.3 | 86.3 | 86.3 | 86 | 90 | 32 | 22.4 | 3.85809 53502 12316 95256 87588 28303 19649 E-56 |

**Table 3.** The values of two-center Coulomb integrals over STO in lined-up coordinate systems

| $n_1/n'_1$ | $l_1/l'_1$ | $m_1/m'_1$ | $\zeta_1/\zeta'_1$ | $n_2/n'_2$ | $l_2/l'_2$ | $m_2/m'_2$ | $\zeta_2/\zeta'_2$ | $R$ | Results |
|---|---|---|---|---|---|---|---|---|---|
| 1/1 | 0/0 | 0/0 | 5.2/5.2 | 2/2 | 0/0 | 0/0 | 4.1/4.1 | 0.2 | 1.82289 25537 50662 68097 06249 99472 18105[a,b,c]<br>1.82289 2554[d,e] |
| 1/2 | 0/1 | 0/0 | 5.2/3.1 | 2/3 | 0/2 | 0/0 | 4.1/2.5 | 0.2 | -2.36064 30209 20063 71569 41492 47834 51963 E-02[a,b]<br>-2.36064 3021 E-02[d,e] |
| 2/1 | 1/0 | 1/0 | 4.0/5.2 | 2/2 | 1/0 | 1/0 | 3.1/4.1 | 0.2 | 2.03568 85382 24252 94658 39569 97218 82382 E-01[a,b]<br>2.03568 8538 E-01[d,e] |
| 1/1 | 0/0 | 0/0 | 5.2/5.2 | 2/2 | 1/1 | -1/-1 | 3.1/3.1 | 8.5 | 1.17392 89654 55745 79366 72606 57106 36806 E-01[a,b,c]<br>1.17392 8965 E-01[d,e] |
| 2/2 | 1/1 | 0/0 | 3.1/3.1 | 4/4 | 2/2 | 2/2 | 0.5/0.5 | 8.5 | 8.75284 77629 56292 02391 86570 66188 70938 E-02[a,b]<br>8.75284 7763 E-02[d,e] |
| 3/3 | 2/2 | -2/-2 | 1.8/1.8 | 2/2 | 0/0 | 0/0 | 4.1/4.1 | 8.5 | 1.15668 97493 85519 57315 49276 08453 94326 E-01[a,b,c]<br>1.15668 9749 E-01[d,e] |
| 4/2 | 3/1 | 0/0 | 3.5/3.1 | 4/4 | 2/3 | 2/2 | 0.5/3.0 | 2.5 | -7.36773 13766 53888 45151 51235 09992 20224 E-05[a,b]<br>-7.36773 1337 E-05[d]<br>-7.36773 13344 146751 E-05[f] |
| 1/1 | 0/0 | 0/0 | 0.99/0.99 | 1/1 | 0/0 | 0/0 | 1.01/1.01 | 0.01 | 6.24916 67058 30088 14983 45518 38351 29936 E-01[a,b,c]<br>6.24916 67102 7413 E-01[g]<br>6.24916 67058 30088 14983 45518 384 E-01[h] |
| 1/1 | 0/0 | 0/0 | 5.2/5.2 | 2/2 | 0/0 | 0/0 | 4.1/4.1 | 100 | 1.00000 00000 00000 00000 00000 00000 00000 E-02[a,b,c]<br>1.00000 00000 00000 00000 00000 000 E-02[h]<br>9.99999 99999 9997 E-03[k] |
| 4/1 | 3/0 | 0/0 | 0.8/0.9 | 3/1 | 2/0 | 0/0 | 1.1/1.2 | 100 | 1.32578 24709 36295 45612 88059 75651 53922 E-10[a,b]<br>1.71159 918548 E-10[g]<br>1.32578 24709 36295 45613 E-10[h] |

[a]Global-adaptive method with Gauss-Kronrod extension

[b]Recurrence relations given for calculation of ${}^{P_1}\mathcal{G}^{N_1,q}$ in Appendices A, B and Refs.[58, 60] with $N_i \in \mathbb{N}^+$

[c]Series representation (Eq. (82)) of ${}^{P_1}\mathcal{G}^{N_1,q}$ with $N_i \in \mathbb{N}^+$ and Ref. [61] for $N_s = 250$

[d]Ref.[71], [e]Ref.[63], [f]Ref.[72], [g]Ref.[73], [h]Ref.[74]

[k]These results obtained in Ref. [71] using the algorithm given in Ref.[75]

**Table 4.** The values of two-center hybrid integrals over STO in lined-up coordinate systems

| $n_1/n'_1$ | $l_1/l'_1$ | $m_1/m'_1$ | $\zeta_1/\zeta'_1$ | $n_2/n'_2$ | $l_2/l'_2$ | $m_2/m'_2$ | $\zeta_2/\zeta'_2$ | $R$ | Results |
|---|---|---|---|---|---|---|---|---|---|
| 1/1 | 0/0 | 0/0 | 5.2/5.2 | 1/2 | 0/0 | 0/0 | 5.2/4.1 | 0.2 | 1.82283 32730 08003 82547 87867 31094 97571[a,b,c]<br>1.82283 3273[d,e] |
| 1/2 | 0/1 | 0/0 | 5.2/3.1 | 2/3 | 1/2 | 1/1 | 4.0/3.0 | 0.2 | -6.86828 29183 60912 21475 49388 55052 76624 E-02[a,b]<br>-6.86828 2918 E-02[d,e] |
| 2/1 | 0/0 | 0/0 | 1.0/1.5 | 1/2 | 0/0 | 0/0 | 1.0/1.5 | 0.5 | 3.52830 59069 42601 45585 21523 21676 29343 E-01[a,b,c]<br>3.52830 59069 426004 E-01[f] |
| 2/2 | 1/1 | 0/1 | 3.1/4.0 | 2/3 | 1/1 | 0/1 | 3.1/1.5 | 2.5 | 5.09491 95301 70106 78339 51218 85823 60340 E-02[a,b]<br>5.09491 9530 E-02[d,e] |
| 4/2 | 3/1 | 0/0 | 3.5/3.1 | 1/2 | 0/0 | 0/0 | 5.2/4.1 | 2.5 | 1.45527 74805 70430 59391 69198 03337 30776 E-04[a,b]<br>1.45527 74805 70430 59398 10389 43559 98442 E-04[c]<br>1.45527 7481 E-04[d,e] |
| 2/2 | 1/1 | 1/1 | 4.0/4.0 | 1/3 | 0/2 | 0/0 | 5.2/5.2 | 8.5 | 6.52569 93988 77690 25939 54456 64202 47822 E-07[a,b]<br>6.52569 9424 E-07[d,e] |
| 2/2 | 1/1 | 0/0 | 3.1/3.1 | 4/3 | 3/2 | 0/0 | 3.5/2.5 | 8.5 | 1.08708 58144 55211 74102 89705 11940 46919 E-05[a,b]<br>1.08708 5814 E-05[d,e] |
| 2/2 | 1/1 | 0/0 | 5.2/5.2 | 2/2 | 0/1 | 0/0 | 5.2/4.1 | 0.3 | 8.99999 85103 06214 17316 79580 94484 15397 E-01[a,b]<br>8.99999 85102 379665 E-01[f] |
| 10/1 | 2/0 | 0/0 | 0.2/5.2 | 4/3 | 3/2 | 1/1 | 2.6/3.0 | 8.5 | 2.36533 58321 44220 60856 72486 90851 06350 E-20[a,b]<br>2.36533 5832 E-20[d]<br>2.36533 61038 280399 E-20[f] |

[a]Global-adaptive method with Gauss-Kronrod extension

[b]Recurrence relations given for calculation of ${}^{P_1}\mathcal{G}^{N_1,q}$ in Appendices A, B and Refs.[58, 60] with $N_i \in \mathbb{N}^+$

[c]Series representation (Eq. (81)) of ${}^{P_1}\mathcal{G}^{N_1,q}$ with $N_i \in \mathbb{N}^+$ and Ref. [61] for $N_s = 250$

[d]Ref.[71]

[e]Ref.[63]

[f]Ref.[72]

**Table 5.** The comparison of methods of computing two-center Coulomb integrals over NSTO in lined-up coordinate systems

| $n_1/n'_1$ | $l_1/l'_1$ | $m_1/m'_1$ | $\zeta_1/\zeta'_1$ | $n_2/n'_2$ | $l_2/l'_2$ | $m_2/m'_2$ | $\zeta_2/\zeta'_2$ | $R$ | Results |
|---|---|---|---|---|---|---|---|---|---|
| 1.1/1.1 | 0/0 | 0/0 | 5.2/5.2 | 2.1/2.1 | 0/0 | 0/0 | 4.1/4.1 | 2.0 | **4.99960 44305 09269 74512 47068** E-01[a]<br>**4.99961** 06137 47474 92577 78874 E-01[b](30)<br>**4.99962** 31580 95475 98094 12185 E-01[b](20)<br>**4.99964** 58052 30194 78329 92300 E-01[b](15)<br>**4.99965** 43225 14863 58247 50368 E-01[b](14)<br>**4.99966** 53763 19254 96182 61413 E-01[b](13)<br>**4.99968** 04424 50836 84467 82256 E-01[b](12)<br>**4.99960 42903** 49048 E-01[c] |
| 1.1/1.1 | 0/0 | 0/0 | 5.2/5.2 | 2.1/2.1 | 0/0 | 0/0 | 4.1/4.1 | 0.2 | **1.74489 32510 67943 65295 27064** E-00[a]<br>**1.74489 4**5172 69239 32119 36369 E-00[b](30)<br>**1.74489 7**0834 71446 60289 72755 E-00[b](20)<br>**1.74490 2**0112 52630 59570 82609 E-00[b](15)<br>**1.74490 4**2617 31309 91262 90263 E-00[b](14)<br>**1.74490 7**4806 43795 04387 74798 E-00[b](13)<br>**1.74491 2**0562 80159 01409 20517 E-00[b](12)<br>**1.74489 32510 679**391 E-00[c] |

[a]Global-adaptive method with Gauss-Kronrod extension

[b]One-center expansion method given in Ref.[18]. The values in parentheses are upper limit of summation $(N_e)$

[c]Cuba numerical integration algorithm

**Table 6.** The values of two-center Coulomb integrals over NSTO in lined-up coordinate systems

| $n_1/n'_1$ | $l_1/l'_1$ | $m_1/m'_1$ | $\zeta_1/\zeta'_1$ | $n_2/n'_2$ | $l_2/l'_2$ | $m_2/m'_2$ | $\zeta_2/\zeta'_2$ | $R$ | Results |
|---|---|---|---|---|---|---|---|---|---|
| 1.1/1.1 | 0/0 | 0/0 | 3.3/7.5 | 2.1/2.1 | 0/0 | 0/0 | 5.2/4.1 | 2.0 | **3.70777 93430 04351 88597 41401** E-01<br>**3.70777 9**1146 97699 E-01c |
| 1.1/1.1 | 0/0 | 0/0 | 3.3/7.5 | 2.1/2.1 | 0/0 | 0/0 | 5.2/4.1 | 0.2 | **1.43116 03370 07045 41502 21987** E-00<br>**1.43116 03370 070**363 E-00c |
| 1.5/2.5 | 0/0 | 0/0 | 5.2/5.2 | 2.2/1.3 | 0/0 | 0/0 | 4.1/4.1 | 2.0 | **4.03973 39319 31391 43106 54998** E-01<br>**4.03973 390**29 016377 E-01c |
| 1.5/2.5 | 0/0 | 0/0 | 5.2/5.2 | 2.2/1.3 | 0/0 | 0/0 | 4.1/4.1 | 0.2 | **1.37950 96732 83485 29390 58534** E-00<br>**1.37950 96732 834**748 E-00c |
| 3.3/1.2 | 0/0 | 0/0 | 3.5/4.5 | 1.3/3.5 | 0/0 | 0/0 | 5.3/5.2 | 2.0 | **1.43069 98608 06165 34609 69226** E-01<br>**1.43069 9**7832 695897 E-01c |
| 1.3/2.2 | 0/1 | 0/0 | 5.2/3.1 | 2.1/3.3 | 0/2 | 0/0 | 4.1/2.5 | 0.2 | **-2.20621 95808 46121 42004 26870** E-02<br>**-2.20621 95808 4612**94 E-02c |
| 2.3/2.5 | 0/1 | 0/0 | 5.2/3.1 | 2.5/3.5 | 0/2 | 0/0 | 4.1/2.5 | 0.2 | **-2.43223 06409 75118 67189 49036** E-02<br>**-2.43223 06409 7511**7 E-02c |
| 2.2/1.4 | 1/0 | 1/0 | 4.0/5.2 | 2.4/2.2 | 1/0 | 1/0 | 3.1/4.1 | 0.2 | **2.02231 33338 48645 17745 17104** E-01<br>**2.02231 33338 486**546 E-01c |
| 1.7/1.7 | 0/0 | 0/0 | 5.2/5.2 | 2.5/2.5 | 1/1 | -1/-1 | 3.1/3.1 | 8.5 | **1.17291 23163 43926 46406 72110** E-01<br>**1.1729**0 70005 783289 E-01c |
| 2.6/2.4 | 1/0 | 1/0 | 4.0/5.2 | 1.4/1.2 | 1/0 | 1/0 | 3.1/4.1 | 0.2 | **2.73167 34925 29429 58648 30570** E-01<br>**2.73167 34925 29**6465 E-01c |
| 3.1/3.1 | 2/2 | -2/-2 | 1.8/1.8 | 2.3/2.5 | 0/0 | 0/0 | 4.1/4.1 | 8.5 | **1.15128 95407 69354 45822 45832** E-01<br>**1.15128** 19943 95839 E-01c |

[a]Global-adaptive method with Gauss-Kronrod extension

[b]Cuba numerical integration algorithm

**Table 7.** The values of two-center hybrid integrals over NSTO in lined-up coordinate systems

| $n_1/n'_1$ | $l_1/l'_1$ | $m_1/m'_1$ | $\zeta_1/\zeta'_1$ | $n_2/n'_2$ | $l_2/l'_2$ | $m_2/m'_2$ | $\zeta_2/\zeta'_2$ | $R$ | Results |
|---|---|---|---|---|---|---|---|---|---|
| 1.1/1.1 | 0/0 | 0/0 | 3.3/7.5 | 2.1/2.1 | 0/0 | 0/0 | 5.2/4.1 | 2.0 | **3.34570 65033 29520 49016 78182** E-02<br>**3.34570** 49544 71777 E-02 |
| 1.1/1.1 | 0/0 | 0/0 | 3.3/7.5 | 2.1/2.1 | 0/0 | 0/0 | 5.2/4.1 | 0.2 | **1.42056 86050 09755 96650 10503** E-00<br>**1.42056 86050 097**461 E-00 |
| 1.5/2.5 | 0/0 | 0/0 | 5.2/5.2 | 2.2/1.3 | 0/0 | 0/0 | 4.1/4.1 | 2.5 | **5.82088 47419 91349 23796 94930** E-03<br>**5.82087** 59642 53641 E-03 |
| 1.5/2.5 | 0/0 | 0/0 | 5.2/5.2 | 2.2/1.3 | 0/0 | 0/0 | 4.1/4.1 | 0.25 | **1.31612 20582 17586 69403 48337** E-00<br>**1.31612 20582 175**52 E-00 |
| 2.3/2.3 | 1/1 | 0/0 | 3.1/4.0 | 2.1/3.5 | 0/1 | 0/0 | 3.1/1.5 | 2.5 | **3.98086 07407 55049 65876 23549** E-01<br>**3.98086 074**72 57121 E-01 |
| 2.3/2.3 | 1/1 | 0/0 | 3.1/4.0 | 2.1/3.5 | 0/1 | 0/0 | 3.1/1.5 | 0.25 | **9.62984 79396 02364 97003 91963** E-02<br>**9.62984 79396 02**41 E-02 |
| 2.3/2.3 | 1/1 | 0/1 | 3.1/4.0 | 2.1/3.5 | 1/1 | 0/1 | 3.1/1.5 | 2.5 | **5.46328 06938 93362 73357 23212** E-03<br>**5.4632**9 74885 12861 E-03 |
| 2.3/2.3 | 1/1 | 0/1 | 3.1/4.0 | 2.1/3.5 | 1/1 | 0/1 | 3.1/1.5 | 0.25 | **1.16785 23072 08658 28310 17007** E-02<br>**1.16785 23072 08**9619 E-02 |
| 2.1/2.2 | 1/1 | 0/1 | 3.1/4.0 | 2.3/3.4 | 1/1 | 0/1 | 3.1/1.5 | 0.85 | **1.19610 29417 94992 00041 70305** E-02<br>**1.19610 2941**8 696619 E-02 |
| 2.1/2.2 | 1/1 | 0/1 | 3.1/4.0 | 2.3/3.4 | 1/1 | 0/1 | 3.1/1.5 | 8.5 | **6.21247 25221 86688 36073 21675** E-06<br>**6**.17573 31239 59217 E-06 |

[a]Global-adaptive method with Gauss-Kronrod extension
[b]Cuba numerical integration algorithm

**Figure 1.** The comparison of numerical and series representation calculations of ${}^{P_1}\mathcal{G}^{N_1,q}$ auxiliary functions depending on $p_1$

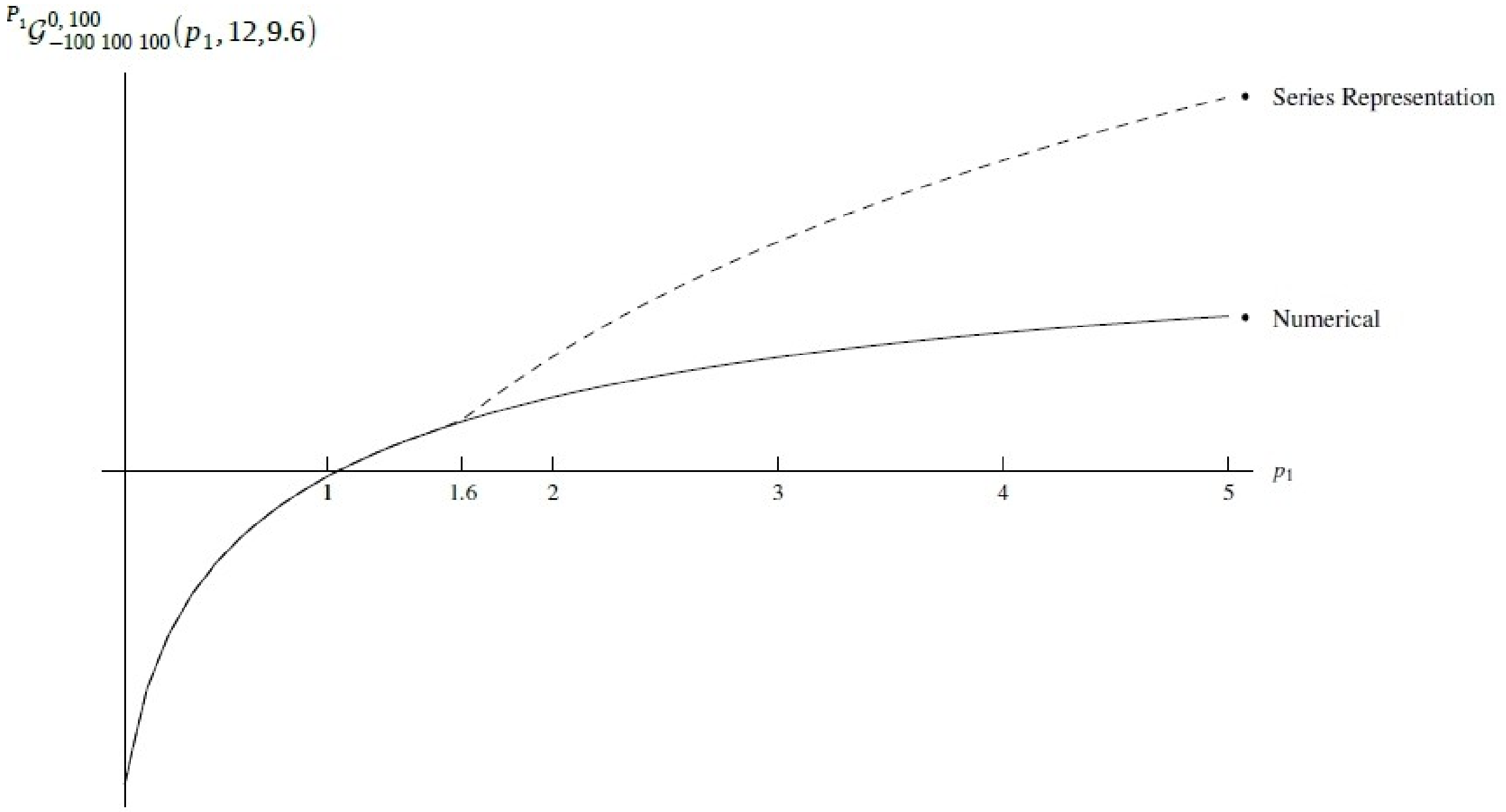

**Figure 2.** The comparison for series representation calculations of ${}^{P_1}\mathcal{G}^{N_1,q}$ auxiliary functions depending on $N_s$ with different values of $p_1$

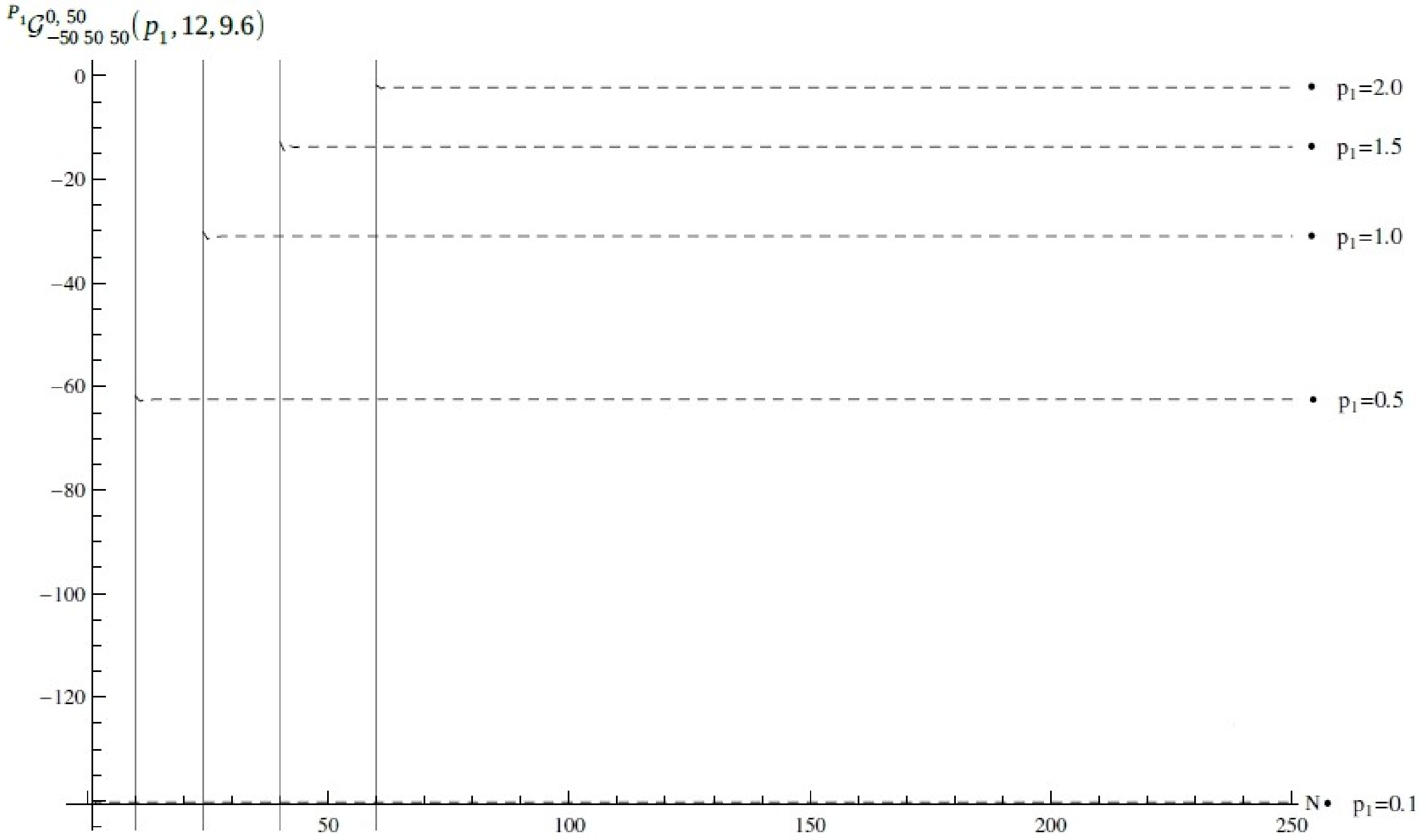